\documentclass[twocolumn,aps,prl,reprint,noshowpacs,letterpaper,floatfix]%
{revtex4-1}
\usepackage{amsmath,amssymb}
\usepackage{txfonts}
\usepackage{bm}
\usepackage{textcomp}
\usepackage[pdftex]{graphicx}
\usepackage{color}
\usepackage{braket}

\begin{document}

\title{Unusual Nonmagnetic Ordered State in CeCoSi Revealed by $^{59}$Co-NMR and NQR Measurements}


\author{Masahiro Manago}
\email{manago@riko.shimane-u.ac.jp}
\altaffiliation{Present address: Department of Physics and Materials Science,
Graduate School of Natural Science and Technology, Shimane University, Matsue, Shimane, Japan.}
\author{Hisashi Kotegawa}
\author{Hideki Tou}
\author{Hisatomo Harima}
\affiliation{Department of Physics, Kobe University, Kobe, Hyogo 657-8501,
Japan}

\author{Hiroshi Tanida}
\affiliation{Liberal Arts and Sciences, Toyama Prefectural University,
Imizu, Toyama 939-0398, Japan}

\begin{abstract}
    We performed $^{59}$Co nuclear magnetic and quadrupole resonance
    (NMR and NQR) measurements under pressure on a single-crystalline CeCoSi,
    which undergoes an unresolved phase transition at $T_0$.
    The NQR spectra clearly showed that the phase transition at $T_0$ is nonmagnetic,
    but any symmetry lowering at the Co site was not seen irrespective of the feature of second-order phase transition.
    By contrast, the NMR spectra were split by the induced magnetic field perpendicular to the external magnetic field.
    These results show that the phase below $T_0$ is not a simple paramagnetic state but is most likely electric multipolar ordered state of Ce $4f$ electrons.
    The development of the Kondo effect by applying pressure
    is thought to be crucial to stabilize this state and to show novel features beyond commonality of tetragonal Ce-based systems.
\end{abstract}

\maketitle

Physical properties of solids are influenced intricately by degrees of freedom of electrons,
that is, charge, spin, and orbital,
through their mutual couplings and many-body interactions of electrons.
Phase transition characterized by
an order parameter is one of the demonstrations of their influence;
therefore, it is a fundamental yet profound phenomenon.
Rich interactions in solids induce various types of phase transitions;
hence, they are attractive in condensed matter physics.

Such diversity yields at times a mysterious ordered state.
A well-known example is the ``hidden order'' state in the $5f$-electron system
URu$_2$Si$_2$ \cite{RevModPhys.83.1301,J.Phys.32.143002}.
It is a second-order phase transition with a symmetry reduction;
however, the crystallographic symmetry of the ordered state is still controversial
\cite{Science.331.439,Philos.Mag.94.3691,Nat.Commun.6.6425,PhysRevLett.124.257601}.
Several researchers have attempted unmasking the order parameter
in URu$_2$Si$_2$ for three decades because it is just a fundamental question
in condensed matter physics; the mechanism through which degrees of freedom of electrons
can affect physical properties of solids.

The $4f$-electrons system CeCoSi is a potential example exhibiting an extraordinary order
parameter.
It crystallizes in the tetragonal $P4/nmm$ ($D_{4h}^{7}$, No.~129)
space-group symmetry with the CeFeSi-type structure \cite{Bodak1970}.
The spacial inversion symmetry is locally absent at the Ce site,
whereas the crystal structure has the global inversion symmetry.
The crystal electric field (CEF) ground state has been reported as the
$\Gamma_7$ ($\mp 0.306\ket{\pm 5/2} \pm 0.95\ket{\mp 3/2}$) Kramers doublet,
and the first excited state is separated from it
by $\sim 100$ K \cite{PhysRevB.101.214426}.
It has been established that an antiferromagnetic (AFM) transition occurs at
the Néel temperature $T_{\textrm{N}} = 9.4$ K.
The transition is of second-order \cite{PhysRevB.70.174408},
and the Ce moments with sizes of $m_{\textrm{Ce}} \sim 0.37(6) \mu_{\textrm{B}}$ are
aligned along the $[100]$ axis with a $\bm{q} = \bm{0}$ structure, as revealed
by a neutron scattering study \cite{PhysRevB.101.214426}.
Another phase, which is a matter of interest,
has been initially reported to emerge under pressure below
$\sim 40$ K at $\sim 1.5$ GPa \cite{PhysRevB.88.155137}.
The first unresolved issue of this phase is its intrinsic pressure phase diagram.
Contrary to some reports \cite{PhysRevB.88.155137,PhysRevB.101.214426},
a study on single-crystalline samples proposed that this
``pressure-induced ordered phase'' already exists at ambient pressure below
$T_0 = 12$ K \cite{JPSJ.88.054716}.
The phase below $T_0$ at lower pressures seems to continuously connect to the pressure-induced phase
\cite{JPSJ.87.023705,JPSJ.88.054716}.
However, the anomaly at $T_0$ at ambient pressure is not sufficiently large to convince a phase transition.
Therefore, microscopic measurements are desired to reveal whether the phase transition at $T_0$
is intrinsic or not at ambient pressure.
The second unresolved issue is the order parameter below $T_0$.
It seems different from  usual AFM states, as deduced from
an enhancement by the magnetic field \cite{PhysRevB.88.155137,JPSJ.88.054716}.
There have been various suggestions for the origin of this phase, as follows:
the spin-density-wave order of Co $3d$ electrons, metaorbital transition,
and antiferroquadrupolar (AFQ) ordering \cite{PhysRevB.88.155137,JPSJ.87.023705,JPSJ.88.054716};
however, the order parameter remains unresolved.

In this Letter, we present the results of
$^{59}$Co nuclear magnetic and quadrupole resonance (NMR and NQR) measurements
on high-quality single-crystalline CeCoSi samples to clarify the above-mentioned issues.
The combined NMR and NQR results revealed the intrinsic phase diagram, where the objective phase was present even at ambient pressure.
In the ordered state, there is no clear indication of symmetry reduction at the Co site under zero field,
whereas the NMR spectra split below $T_0$ when the magnetic field tilted from the [100] axis was applied.
This NMR anomaly can be interpreted by the emergence of the induced magnetic
field perpendicular to the external field.
Present results suggest an unusual nonmagnetic ordered state that is most likely an electric multipole ordered state in CeCoSi.

Plate-shaped single-crystalline CeCoSi samples
($\sim 3 \times 4 \times 0.5 $ mm$^{3}$) were grown using the Ce/Co eutectic
flux method, as described in Ref.~\onlinecite{JPSJ.88.054716}.
The $^{59}$Co ($I=7/2$) NMR and NQR measurements were performed in the
temperature range 1.4--300 K
using a standard spin-echo method.
The NMR spectra were obtained under a field of 1 T near the $[100]$ direction,
and its angle was deduced from the frequencies of the spectra above $T_0$.
The local symmetry of the Co site in the $P4/nmm$ space group
is $\bar{4}m2$ with four-fold improper rotational symmetry.
Hydrostatic pressure was applied up to 2.35 GPa on another sample using a
piston-cylinder-type cell with Daphne 7474 as a pressure-transmitting medium.
The pressure was determined from the superconducting transition temperature
of a Pb sample inside the cell.
A LaCoSi sample consisting of single-crystalline pieces was also measured by NQR at $P=0$
as a reference system.

\begin{figure}
    \centering
    \includegraphics{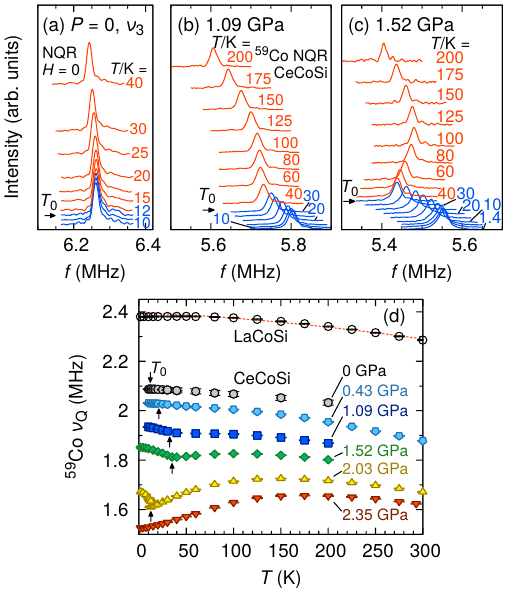}
    \caption{\label{fig:nqr-fsp}(Color online)
        (a--c) Temperature dependence of the $^{59}$Co NQR spectra of CeCoSi at
        $\nu_3 \equiv 3\nu_\textrm{Q}$ site without a field under a pressure of
        (a) 0, (b) 1.09, and (c) 1.52 GPa.
        The spectra at each temperature are shifted vertically.
        (d) Temperature and pressure dependence of $\nu_{\textrm{Q}}$
        of CeCoSi and that of LaCoSi at ambient pressure.
        The vertical arrows indicate $T_0$ determined by NMR measurements, which detected it even at ambient pressure.
        The dashed line for LaCoSi indicates the conventional temperature dependence of $\nu_{\textrm{Q}}$.
    }
\end{figure}

The NQR spectra are sensitive to both the magnetic and electric anomalies
and are suitable for probing the nature of the ordered state.
Figures \ref{fig:nqr-fsp}(a)--\ref{fig:nqr-fsp}(c) show the temperature dependence of the
NQR spectra above $T_{\textrm{N}}$ at the $\nu_3 \equiv 3\nu_{\textrm{Q}}$
($\pm 5/2 \leftrightarrow \pm 7/2$) site
at 0, 1.09, and 1.52 GPa.
The value of the quadrupole frequency $\nu_{\textrm{Q}}$ was 2.09 MHz at $P=0$ and 20 K\@.
The results on other pressures are shown in the Supplemental Materials \cite{SM}.
The NQR spectra did not split nor broaden, although they showed a shift at $T_0$.
As shown in the temperature dependence of $\nu_{\textrm{Q}}$ in Fig.~\ref{fig:nqr-fsp}(d),
the kink of $\nu_{\textrm{Q}}$ is clearer at higher pressures, and it is invisible at ambient pressure.
The kink is not a necessary condition for the phase transition,
and its presence at ambient pressure is proved by NMR measurements mentioned later.
The absence of splitting or broadening NQR spectra demonstrates that the internal field is absent at the Co sites for
$T_{\textrm{N}}<T<T_0$ in the entire pressure range.
This clearly excludes the magnetic ordering of Co $3d$ moments below $T_0$.
Our results also indicate that all the Co sites remain equivalent below $T_0$
while maintaining a local tetragonal symmetry, that is, a tetragonal crystal structure \cite{SM}.
If one considers a possibility of the magnetic ordering of Ce $4f$ moments,
the only explanation for the NQR result is a cancellation of the internal
magnetic field at the Co site, which is surrounded by four Ce ions.
The transition to the underlying AFM state at $T_\textrm{N}$ is of second-order,
and this AFM state with $\bm{q}=\bm{0}$ does not break the translational
symmetry of the crystal above $T_0$.
Then, the intermediate ordered state in $T_{\textrm{N}} < T < T_0$
should maintain the same translational symmetry, that is, the propagation vector is $\bm{q}=\bm{0}$.
For the $\bm{q}=\bm{0}$ structure,
the internal field is canceled only when the staggered
Ce moment is along the $[001]$ axis \cite{SM}.
This possibility is excluded by the increase of the susceptibility
along this axis below $T_0$ \cite{JPSJ.88.054716}.
Thus, we concluded that the ordered phase below $T_0$ is nonmagnetic with one
tetragonal Co site.
The unusual field response explained later
also excludes the possibility of the magnetic state and yet indicates that
this phase is not a simple paramagnetic state.

The temperature dependence of $\nu_{\textrm{Q}}$ usually obeys a conventional monotonous behavior
$\nu_{\textrm{Q}}(T)=\nu_{\textrm{Q}}(0) (1-\alpha T^{1.5})$
\cite{ZPhysB.24.177} ($\alpha > 0$) owing to the lattice expansion and vibration,
as observed in LaCoSi.
In contrast, the $\nu_{\textrm{Q}}(T)$ in CeCoSi deviates from this behavior
above $T_0$, and the deviation becomes remarkable as the pressure increases.
Such behavior has been observed in some $4f$-electron systems
\cite{JPSJ.56.4113,PhysRevB.96.134506,JPSJ.82.103704,JPSJ.81.124706,JPSCP.3.011046}
and discussed to originate from a CEF splitting, a valence change of the $4f$ ion, or the Kondo effect.
The distinct pressure dependence in CeCoSi indicates that such multiple effects influence $\nu_{\textrm{Q}}$ at the Co site
through the pressure-enhanced hybridization between the $4f$ electron
and conduction electrons.

\begin{figure}
    \centering
    \includegraphics{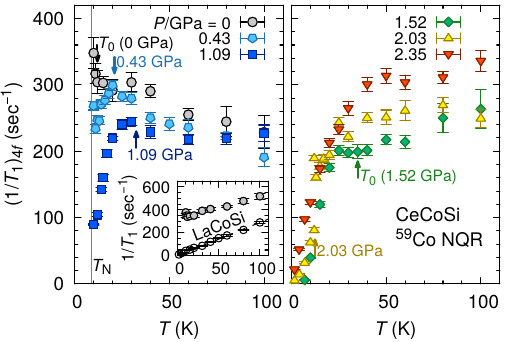}
    \caption{\label{fig:t1}(Color online)
        Temperature dependence of $^{59}$Co NQR nuclear spin-lattice relaxation rate $1/T_1$
        of CeCoSi from the Ce-$4f$ electrons at $H=0$
        under several pressures.
        The $4f$ part was obtained by subtracting the $1/T_1$ value of LaCoSi.
        The vertical arrows indicate $T_0$ determined by NMR results.
        Inset (left panel): $1/T_1$ of CeCoSi at ambient pressure
        before subtraction as well as that of LaCoSi.
        The dashed line for LaCoSi shows the Korringa relation
        $1/T_1 \propto T$ in the normal metal.
    }
\end{figure}

Figure \ref{fig:t1} shows the Ce-$4f$ part of the nuclear spin-lattice
relaxation rate $1/T_1$ measured by the NQR at $H=0$.
The $1/T_1$ has a Co $3d$ component, and it was subtracted using the result
of LaCoSi (see the inset of the left panel of Fig.~\ref{fig:t1}).
The $1/T_1$ divided by temperature is shown in Supplemental Materials \cite{SM}.
The $1/T_1$ of LaCoSi shows the weakly-correlated Pauli paramagnet
corresponding to a previous report \cite{J.Alloys.Compd.210.279}.
The absence of phase transition in LaCoSi is consistent with the interpretation that the phase below $T_0$ originates from
Ce $4f$ electrons.
No divergent behavior was detected near $T_0$ in $1/T_1$ in CeCoSi,
indicating the absence of the magnetic critical fluctuations.
A clear drop was found at $T_0$ only at 2.03 GPa.
This may correspond to the gaplike anomaly in the electric resistivity
above $\sim 1.5$ GPa \cite{PhysRevB.88.155137}
and suggest the decrease of the density of states.
Around $T_{\textrm{N}}$,
$1/T_1$ is expected to detect the magnetic fluctuations perpendicular to the $[001]$ axis \cite{SM}.
$1/T_1$ diverged owing to the critical slowing down of the magnetic fluctuations
at ambient pressure (see also Fig.~7 in the Supplemental Materials \cite{SM}).
However, the divergence of $1/T_1$ is suppressed with increasing pressure.
This is not expected with the magnetic structure of $\bm{m} \parallel [100]$.
Two possibilities are considered to interpret this result.
One is that the magnetic moment is tilted along the [001] axis under pressure.
Another is that the magnetic fluctuation is suppressed
as the ordered state below $T_0$ gets stabilized.
The $1/T_1$ significantly above $T_0$ shows the localized Ce $4f$
electrons ($1/T_1 = \textrm{const.}$) at 1.52 GPa or below.
Meanwhile, $1/T_1$ starts to decrease from higher temperatures than $T_0$ at 2.03 and 2.35 GPa.
Such an itinerant behavior owing to coherent Kondo effect in $1/T_1$ is also seen in some heavy-fermion systems
including CeCu$_2$Si$_2$ \cite{PhysRevLett.82.5353,JPSJ.77.123711}.
The $1/T_1$ result, combined with the $\nu_{\textrm{Q}}$  behavior,
indicate an enhancement of the Kondo effect by pressure.

\begin{figure}
    \centering
    \includegraphics{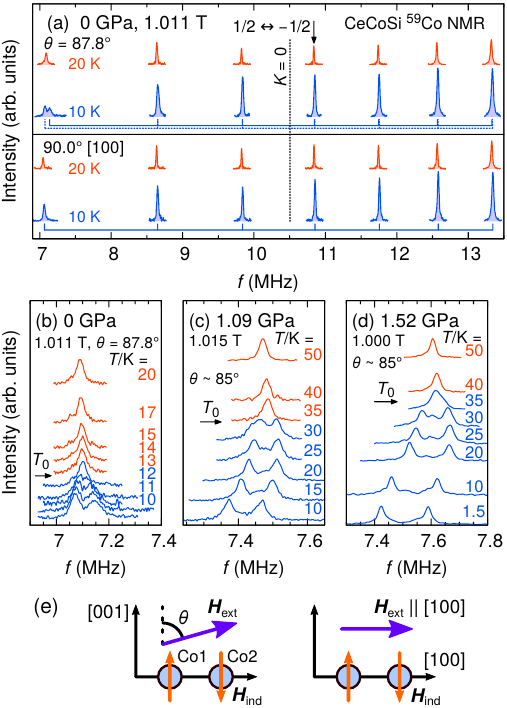}
    \caption{\label{fig:nmr-spectra}(Color online)
    (a) $^{59}$Co NMR spectra of CeCoSi at ambient pressure at 10 and 20 K
    with the field $\mu_0 H = 1.011$ T and angles
    $\theta = 87.8$\textdegree\ and 90\textdegree.
    The blue solid and dashed lines indicate the simulated peak frequencies
    at $T=10$ K assuming the staggered
    field of $\sim \pm 15$ mT along the $[001]$ axis.
    The vertical arrow indicates the central ($1/2 \leftrightarrow -1/2$)
    line, whereas the vertical black dashed line indicates the frequency of $K=0$
    of the central line.
    (b--d) Temperature dependence of the spectra
    under pressures of (b) 0, (c) 1.09, and (d) 1.52 GPa with $\mu_0 H = 1.0$ T
    slightly titled from the $[100]$ axis.
    The spectra are shifted vertically with an offset proportional to the
    temperatures.
    The horizontal arrows indicate $T_0$.
    (e) Schematic of the induced field $\bm{H}_{\textrm{ind}}$
    under the external field $\bm{H}_{\textrm{ext}}$
    at the Co sites below $T_0$.
    Only the $[001]$ components of $\bm{H}_{\textrm{ind}}$
    are drawn for simplicity.
    }
\end{figure}

Further information on the ordered state is provided by the spectra measured under a magnetic field.
Figure \ref{fig:nmr-spectra}(a) shows the full set of the NMR lines at ambient
pressure at 10 K below $T_0 = 12$ K with a field strength of $\mu_0 H = 1.0$ T
in addition to the results of 20 K.
Here, $\theta = 87.8$\textdegree\ and 90\textdegree\ are the field
angles from the $[001]$ to $[100]$ axes.
At 20 K, seven NMR lines were observed owing to the nuclear
quadrupole interaction of $I=7/2$ $^{59}$Co nucleus.
Meanwhile, at 10 K, the third satellite at the lowest frequency
split when the field was tilted from the $[100]$ axis,
clearly indicating the symmetry reduction below $T_0$.
The two-split peaks merged just when $H \parallel [100]$,
which was consistent with the NQR data and suggesting that the crystallographic Co site
remains one site below $T_0$.
Similar split and merging of the spectra were observed when the magnetic field is around $[001]$ axis (not shown).

Figures \ref{fig:nmr-spectra}(b)--\ref{fig:nmr-spectra}(d) show the temperature dependence of the
NMR third satellite line with the field of $\mu_0 H = 1.0$ T
slightly tilted from the $[100]$ axis
under pressures of 0, 1.09, and 1.52 GPa.
The results under other pressures are shown in Supplemental Materials \cite{SM}.
The spectra started to split at $T_0$ at all pressures, and the value of $T_0$
corresponds with the previous reports \cite{JPSJ.87.023705,JPSJ.88.054716}.
This is the first microscopic evidence for the phase transition
at $T_0$ at ambient pressure.
Careful measurements just below $T_0$, especially at ambient pressure,
indicate that the transition is of second-order \cite{SM}.
Three peaks were observed at intermediate temperatures
at 1.52 GPa and other pressures \cite{SM}, although it is unclear whether this
is intrinsic or not.
Here we assume the two-peak structure to be intrinsic because it is also
observed at ambient pressure, where the stress or inhomogeneity owing to pressure is free.

The NMR line split is remarkable at the low-frequency third satellite line and disappears for $H \parallel [100]$.
This situation is reproduced only when the induced magnetic field
$\bm{H}_{\textrm{ind}}$ is perpendicular to the external field, as shown in
Fig.~\ref{fig:nmr-spectra}(e).
The blue solid and dashed lines in Fig.~\ref{fig:nmr-spectra}(a) indicate the simulated
NMR frequencies, where $\mu_0 H_{\textrm{ind},c} \sim \pm 15$ mT along the $[001]$ axis are adopted.
When the external field $\bm{H}_{\textrm{ext}}$ is tilted from $[100]$ to the $[001]$ axis,
the total field at two Co sites, $\bm{H}_{\textrm{ext}}+\bm{H}_{\textrm{ind}}$,
differs from each other, leading to the splitting of the NMR frequency.
The field $\mu_0 H_{\textrm{ind},z} \sim \pm 15$ mT is absent in the NQR spectra,
and thus, this is induced by the external field.
Such an induced field cannot be explained by a simple paramagnetic response of Ce $4f$
magnetic moment \cite{SM}.
Notably, this excludes the possibility that the internal field at the Co site is
canceled out in the AFM alignment of the Ce moments
because tilting of the AFM moments under an external field yields
a similar response to the paramagnetic response without splitting of the NMR line.

\begin{figure}
    \centering
    \includegraphics{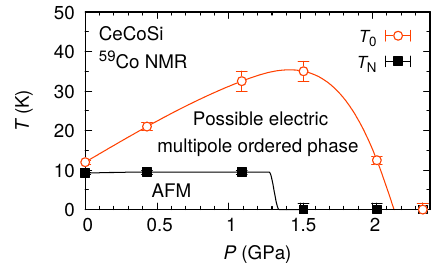}
    \caption{\label{fig:phase-diagram}(Color online)
        Temperature--pressure phase diagram of CeCoSi determined by the
        NMR and NQR results.
        The solid lines are guides to the eye.
        The transitions are of second-order across $T_0$ and $T_{\textrm{N}}$.
        The disappearing pressures of two phases are from
        Ref.~\onlinecite{PhysRevB.88.155137}.
    }
\end{figure}

Figure \ref{fig:phase-diagram} shows the temperature--pressure phase diagram
of CeCoSi determined by our NMR and NQR results.
The phase transition at $T_0$ at ambient pressure was unambiguously confirmed to be intrinsic
from the microscopic point of view.
Our results clearly demonstrate that second-order transitions occur successively across $T_0$ and $T_\textrm{N}$.
The pressure dependence of $T_0$ is reminiscent of the Doniach-type phase diagram,
implying that the $c$--$f$ hybridization assists the stabilization of the ordered state at lower pressures.
Two ordered phases are suppressed with increasing pressure,
accompanied by the development of the Kondo effect,
and the nonmagnetic ground state is realized in the pressure range of 1.4--2.15 GPa.

Our NMR and NQR results limit possible order parameters in CeCoSi as follows:
in $T_{\textrm{N}} < T < T_0$, they exclude magnetically ordered states arising from Ce-$4f$ and Co-$3d$ electrons.
Nonmagnetic ordered states arising from Co-$3d$ electrons, such as a charge density wave
(CDW) or orbital ordering are also unlikely.
If the CDW transition occurs, the spectra will be split into several sites
corresponding to the superlattice formation.
Similarly, if the phase transition lifts the Co-$3d$ orbital degeneracy protected by local tetragonal symmetry,
orthorhombic transition is essential similar to an electronic nematic state in Fe pnictide systems\cite{Nature.486.382}.
The on-site NMR and NQR results did not show the corresponding symmetry reduction.
As an ordered state of nonmagnetic Ce origin, the ``metaorbital'' transition
\cite{JPSJ.79.114717}, across which the occupancy of the Ce-$4f$ electrons
changes steeply, is proposed \cite{PhysRevB.88.155137};
however, it does not correspond with the second-order transition
with a symmetry reduction.
The remaining possibility that should be considered is the contribution
of electric degrees of freedom of
Ce $4f$ electrons; that is, an electric multipole ordering.
The CEF ground state of tetragonal CeCoSi is a Kramers doublet,
but the AFQ ordering is possible if the state including the first-excited level possesses
a large interlevel interaction \cite{JPSJ.89.013703}, which is most likely enhanced by the Kondo effect.
Quadrupolar degrees of freedom involving CEF excited states are also known
in PrOs$_4$Sb$_{12}$ \cite{JPSJ.72.1002,PhysRevB.69.180511} and CeTe \cite{JPSJ.80.023713}.

Before deepening this discussion, we provide a general symmetry
consideration, as discussed for the hidden-order state in URu$_2$Si$_2$
\cite{JPSJ.79.033705,PhysRevB.97.235142}.
Because the transition to the nonmagnetic state across $T_0$ in CeCoSi is second-order,
the space group below $T_0$ must be one of the subgroups
of the room-temperature phase of No.~129 ($P4/nmm$).
Moreover, the underlying AFM state orders to $\bm{q}=\bm{0}$ structure,
preserving the translational symmetry \cite{PhysRevB.101.214426} while
excluding the possibility of superlattice formation above $T_{\textrm{N}}$.
Thus, the possible maximal subgroups are as follows:
No.~59 ($Pmmn$), 85 ($P4/n$), 90 ($P42_12$), 99 ($P4mm$), 113 ($P\bar{4}2_1m$),
and 115 ($P\bar{4}m2$) \cite{InternationalTable,SM}.
Because the Co site remains one site below $T_0$,
the space group with two Co sites, namely, No.~115, is unlikely.
The breaking of the four-fold symmetry was not detected at the Co site,
which contradicts No.~59, 90, and 99 with two-fold symmetry at the Co site.
Therefore, the space group No.~85 or No.~113 is preferable for the state below
$T_0$ in CeCoSi.
An interesting feature is that the symmetry reduction to No.~85 or No.~113
does not require the shift of the atomic positions
\cite{SM,InternationalTable}, that is, it can be satisfied
by symmetrical change of electronic configurations of the Ce ions.
If that is the case, the lattice distortion is triggered by the weak coupling to the electrons,
and the detection of the symmetry change may not be straightforward in X-ray measurements.
This would happen in URu$_2$Si$_2$ \cite{JPSJ.79.033705,PhysRevB.97.235142}.
Moreover, the No.~113 lacks global inversion symmetry;
therefore, experimental methods that can observe the splitting of bands will be effective to confirm this.

A clue to understand the origin of this phase is the unusual field response,
i.e., the induced field perpendicular to the external field.
Such a behavior is reminiscent of an AFQ state \cite{JPSJ.66.1741,JPSJ.66.3005}, as observed in
ferro- or antiferroquadrupolar systems CeB$_6$ \cite{JPSJ.52.728}, PrFe$_4$P$_{12}$
\cite{PhysRevB.71.024424,JPSJ.76.043705}, PrTi$_2$Al$_{20}$ \cite{JPSJ.85.113703},
and NpO$_2$ \cite{PhysRevLett.94.137209}.
Therefore, it is reasonable to consider the ordered state in CeCoSi as an AFQ state.
Along this line of thought, a crucial point is consistency
between the experiment and theoretical suggestion \cite{JPSJ.89.013703}.
In the case of multipolar states, the space groups No.~85 and 113,
suggested by the NQR result, correspond to the
hexa\-deca\-polar and $O_{xy}$-type AFQ states, respectively.
The symmetry of the Ce site reduces to four-fold symmetry without mirror operations ($4..$) in the No.~85
and two-fold symmetry ($2.mm$) in the No.~113 space group \cite{SM}.
Meanwhile, identification of the AFQ order parameter is possible in principle
from the NMR-line splitting by a theoretical work\cite{PhysRevB.102.195147}.
For example, the NMR line can split in the $O_{zx}$-type AFQ state when $H \perp [010]$, except for $H \parallel [100]$ and $H \parallel [001]$.
Therefore, the induced field detected by our NMR suggests the
$O_{zx}$-type AFQ state.
In the case of $O_{xy}$ type,
which is proposed by the zero-field result,
the line splitting does not occur by the component along [001] of the field.
Thus, we need to resolve this discrepancy between zero- and
finite-field results to settle the origin of the ordered state.
A point to be considered is the switching of the order parameter under field,
as in some quadrupole systems such as CeB$_6$ \cite{JPSJ.64.3941} and PrTi$_2$Al$_{20}$ \cite{JPSJ.88.084707}.
Such a signature has not been observed thus far in CeCoSi,
but investigations for the field-induced switching may unravel this issue.
Another point is the lack of local inversion symmetry at the Ce site.
In this case, odd-parity multipoles, such as electric octapole, are active in principle
through the antisymmetric spin-orbit interaction at the Ce site,
but the mechanism through which this effect influences the field response is unknown.
In any case, our results capturing a
peculiarity of the ordered state in CeCoSi will offer theoretical and further experimental challenges
for the complete elucidation of the order parameter.

In conclusion, we have performed single-crystalline NMR and NQR measurements to investigate the unresolved ordered state in CeCoSi.
The absence of any signature of symmetry reduction in the $^{59}$Co-NQR spectra indicates that the ordered phase below $T_0$ is
nonmagnetic and originates from Ce $4f$ electrons.
An unusual field response, proved by the NMR-line splitting, characterizes
the extraordinary ordered state, indicating that it is most likely
the electric multipole state.
Our results show that CeCoSi undergoes two types of second-order phase transitions
with different symmetry lowerings, and the nonmagnetic phase can be the ground state above $\sim 1.4$ GPa.
It is interesting to consider why CeCoSi differs from other non-cubic systems.
As an idea, we expect that parity mixing by absence of the local inversion symmetry at the Ce site may be a key factor to induce this peculiar behavior in CeCoSi.
In any case, CeCoSi is a novel example offering profound
physics originating from the $4f^1$ state.

\begin{acknowledgments}
    The authors thank M. Yatsushiro, S. Hayami, H. Hidaka, Y. Tokunaga,
    K. Ishida, and Y. Kuramoto for their insightful discussions.
    This work was supported by a Grant-in-Aid for Scientific Research on Innovative
    Areas ``J-Physics'' (Grants No.~15H05882, No.~15H05885, No.~JP18H04320, and No.~JP18H04321),
    and a Grant-in-Aid for JSPS Research Fellow (Grant No.~JP19J00336) from JSPS.
\end{acknowledgments}

\bibliography{bibliography}

\end{document}